+

\documentclass[twocolumn,showpacs,prl,superscriptaddress]{revtex4}

\usepackage[dvips]{graphicx}

\begin{document}


\title{Uphill solitary waves in granular flows}


\author{E. Mart{\'i}nez, C. P{\'e}rez-Penichet, O. Sotolongo-Costa}
\affiliation{``Henri Poincar{\'e}" Group of Complex Systems, Physics
Faculty, University of Havana, 10400 Havana, Cuba }

\author{O. Ramos, K. J. M{\aa}l{\o}y}
\affiliation{Physics Department, University of Oslo, Blindern,
N-0316 Oslo, Norway }

\author{S. Douady}

\affiliation{Laboratoire Matière et Systèmes Complexes (MSC),
Université Paris 7 - Denis Diderot / CNRS, CC 7056 75251 Paris
Cedex 05 - France}

\author{E. Altshuler}
\email[]{ea@infomed.sld.cu}
\affiliation{``Henri Poincar{\'e}" Group of Complex Systems and
Superconductivity Laboratory, Physics Faculty-IMRE, University of
Havana, 10400 Havana, Cuba}


\date{\today}

\begin{abstract}

We have experimentally observed a new phenomenon in the surface
flow of a granular material. A heap is constructed by injecting
sand between two vertical glass plates separated by a distance
much larger than the average grain size, with an open boundary. As
the heap reaches the open boundary, ``soliton-like" fluctuations
appear on the flowing layer, and move ``up the hill" (i.e.,
against the direction of the flow). We explain the phenomenon in
the context of stop-and-go traffic models, and show that
soliton-like behavior is allowed within a Saint-Venant description
for the granular flow.


\end{abstract}

\pacs{45.70.-n,45.70Mg,45.70.Vn,81.05Rm,89.75.-k}
\keywords{Granular flows, Granular matter, Soft condensed matter}

\maketitle


The rich dynamics of granular matter --studied for centuries by
engineers-- has attracted much attention from the Physics community
since the early 1990's \cite{Jaeger1996,DeGennes1999,Kadanoff1999}.
Granular flows, for example, have concentrated intense interest, due
to their relevance to natural avalanches and industrial processes,
and also because they make liquid-like and solid-like behaviors
coexist. They have been theoretically described based on the
existence of two phases: the rolling (or flowing) one, and the
static one. Such idea has been casted into {\it ad hoc}
phenomenological equations \cite{BCRE1994,deGennes1995,Aradian1999},
into a Saint-Venant hydrodynamic approach conveniently modified to
take into account the particularities of granular matter
\cite{Douady1999,Andreotti2002}, and eventually by defining an order
parameter characterizing the local state of the system
\cite{Aranson2001,Aranson2002}. Finally, granular flows have been
described by ``microscopic" equations based on the newtonian motion
of individual grains submitted to gravity, shocks, and trapping
events \cite{Quartier2000,Andreotti2001}.

In particular, researchers have studied both experimentally and
theoretically granular flows on inclined planes and tubes, and
granular heaps, finding a whole jungle of patterns,
spatio-temporal structures, and other nontrivial phenomena, such
as fingering \cite{Pouliquen1997}, avalanches extending both
downhill and uphill \cite{Daerr1999}, ``logitudinal vortices"
\cite{Forterre2001,Forterre2002}, ``bubbling flows"
\cite{Flekkoy2001,Gendron2001},``revolving rivers"
\cite{Altshuler2003} and even ``singing" dunes \cite{Douady2005}.

So it looks very unlikely that one can find further unexplored
phenomena in experiments as simple as pouring sand on a heap with an
open boundary. However, we report here the existence of soliton-like
waves moving uphill in such experiments, i.e., the spontaneous
appearance of bump-like instabilities in the flowing layer that
propagate uphill, contrary to the flow of sand.


We used sand with a high content of silicon oxide and an average
grain size of $100 \mu m$ from Santa Teresa (Pinar del R\'{\i}o,
Cuba) \cite{Altshuler2003}. The sand was poured into a cell
consisting in a horizontal base and a vertical wall, sandwiched
between two square glass plates with inner surfaces separated by a
distance $w$ in the range from $0.3 cm$ to $3 cm$ (Fig. 1). The
lengths of the base and the vertical wall were approximately
$d\approx 23 cm$. The sand was poured vertically into the cell
near the vertical wall using tiny funnels with several hole
diameters, in order to obtain different flux values. As the sand
was poured into the cell, a heap grew until it reached the open
boundary, where the grains were allowed to fall freely (Fig. 1).
Digital videos were taken using a High Speed Video Camera Photron
FASTCAM Ultima-APX model 120K in the range from $50 fps$ to $4000
fps$, with a resolution of $1024 \times 1024 $ pixels.


\begin{figure}
\includegraphics[height=1.8in, width=2.0in]{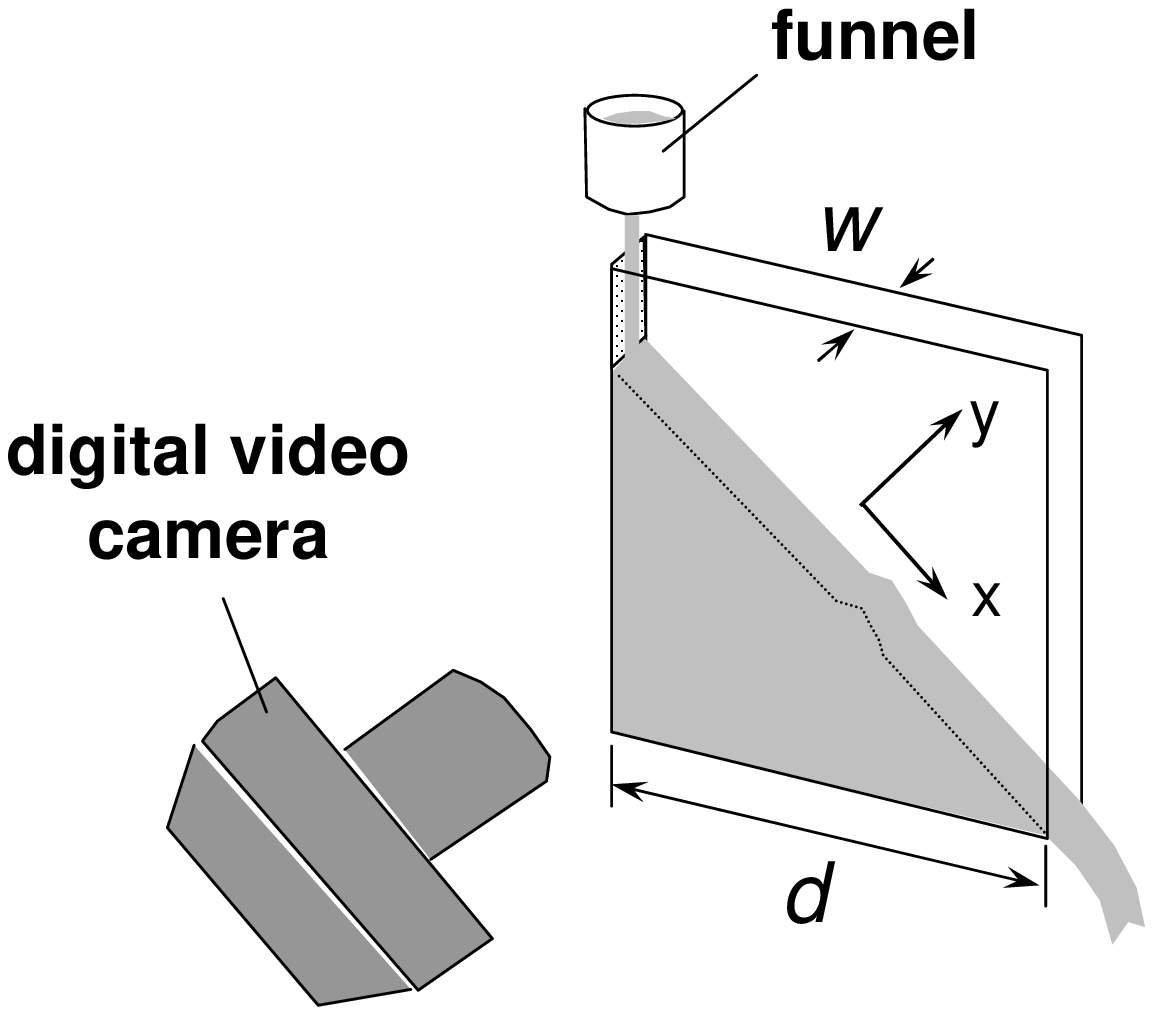}
\caption{\label{fig:f1} Simplified diagram of the experimental
setup}
\end{figure}

\begin{figure*}
\includegraphics[height=2.2in, width=7in]{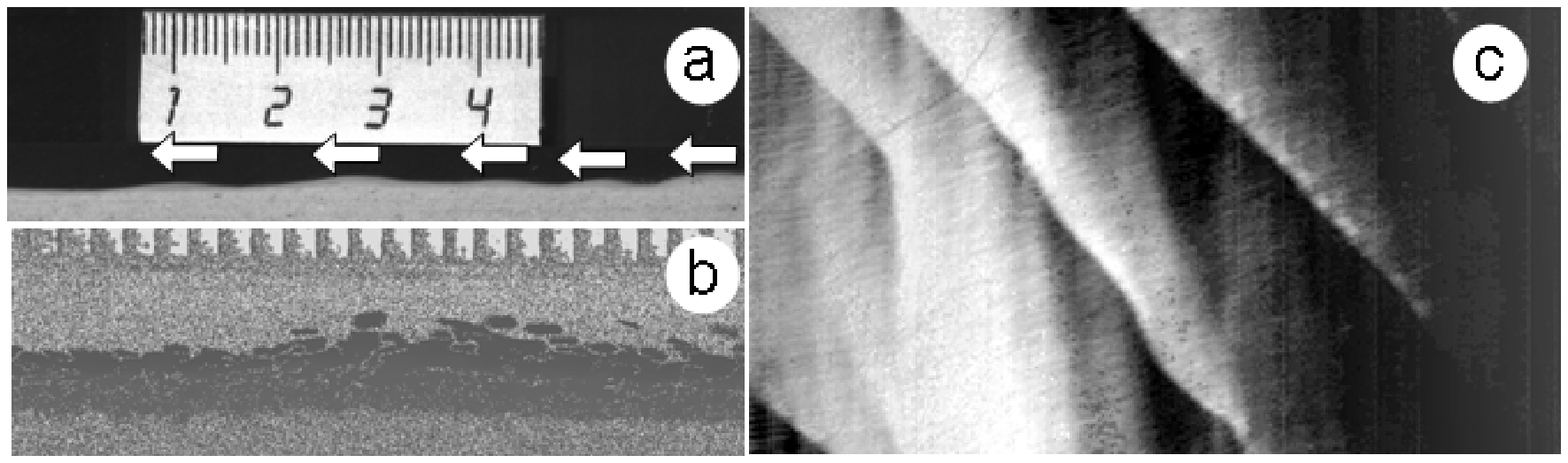}
\caption{\label{fig:f2} Basic experimental findings. (a) Uphill
bumps moving in tandem, identified by arrows. (b) Difference
between two pictures of a single bump, separated by $50 ms$. In
both pictures, the smallest ticks in the scales correspond to $1
mm$. (c) Spatial-temporal diagram of bump dynamics, where $x$
grows to the right and $t$ grows upward, spanning $20 cm$ and $2
s$, respectively.}
\end{figure*}

Fig. 2 contain images from experiments with $w=0.7 cm$ and $F= 0.6
cm^3/s$. Fig. 2 (a) is a picture taken from a video recorded at
$50 fps$, and show a closeup of a section of the surface at the
central region of the heap, where our basic findings are easily
identified: ``bump-shaped" instabilities appearing at random
places of the surface move ``up the hill", as indicated by the
white arrows. These bumps could travel either alone or in tandem.
The bumps maintain their "identity" within variable times, which
can reach more than $10$ seconds. After that, they flatten out. We
have also observed that, if two of them move at speeds different
enough to interfere with each other, they can be still identified
after the interaction. However, these are rare events, and we have
no video record of them. All in all, our observations hint at a
soliton-like behavior in the observed instabilities.

Fig. 2(b) shows the difference between two pictures from a single
bump taken from a video recorded at $500 fps$, which are separated
$50 ms$ from each other. The darker horizontal band allows to
visualize the downhill motion. Careful examination of picture 2(b)
allows to separate an approximately $1 mm$-depth band of ``flying"
grains, and an approximately $1 mm$ depth layer of ``flowing"
grains underneath. The height of the perturbation above the
unperturbed stream is difficult to measure, but it can be roughly
estimated as $1/5$ of the flowing layer.

Fig. 2(c) shows a spatial-temporal diagram of the free surface. A
horizontal line of the video record was taken just at the
(average) free surface, so lower parts appear black, and higher
parts appear white. From the picture it can clearly be seen four
shock-waves, passing suddenly from black to white, moving upward
with a well defined velocity (even if it fluctuates a little on
the right). On the left the appearance of a new soliton can be
seen, first increasing in height before starting suddenly to
propagate.

The average speed of the uphill motion of the bumps as a function
of the input flux, ${\it F}$ was estimated from the analysis of
the motion of several instabilities using pictures analogous to
those used to obtain Fig 2(b). Fig. 3 presents the dependence of
the average uphill speed versus input flux. The solid line
corresponds to $v_{up} \sim F^{1/4}$, which will be discussed
below.

Fig. 4 shows a sequence of pictures of a bump separated by
intervals of $0.125 s$, extracted from a video taken at 4000 fps.
Careful inspection of the video associated to this sequence of
images reveals the mechanism of movement of the instability at the
single grain level. The grains involved in the flowing layer in
picture (a) show two types of behavior: while most of the grains
keep flowing from left to right at an average speed $v_{flow} \sim
10cm/s$ as the bump passes underneath from right to left , a
fraction of the grains at the lower part of the flowing layer show
a different behavior. The white circle in Fig. 4(a), represents
one of those grains, which is moving from left to right. From 4(b)
to 4(d), the grain has incorporated into the static layer
(deposition), while the bump is passing from right to left above
it. In Fig. 4(e), the grain has reincorporated into the flowing
layer (erosion), moving again from left to right.

\begin{figure}
\includegraphics[height=2in, width=2.7in]{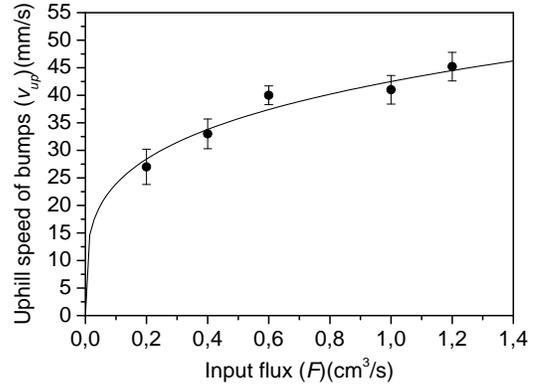}
\caption{\label{fig:f3} Uphill speeds of bumps measured as a
function of input flux, for $w=0.7mm$}
\end{figure}

Exchange of particles between the static and fluid phases has been
considered in several models of surface flows since the 1990's
\cite{BCRE1994}. We believe that our bumps form when a fluctuation
implies an extra deposition of grains from the flowing layer on
the static one, which then propagates backwards through an
``stop-and-go" mechanism observed in traffic dynamics
\cite{Helbing2001}. However, in our case the usual traffic models
should be modified to take into account the free surface. A way to
do it could be a two-lane system consisting in a fast lane where
cars flow steadily, and a slow lane where cars stop and go,
producing a backward wave (or train of waves) within the slow
lane, well described in the literature \cite{Helbing2001}.

\begin{figure}
\includegraphics[height=3in, width=2.7in]{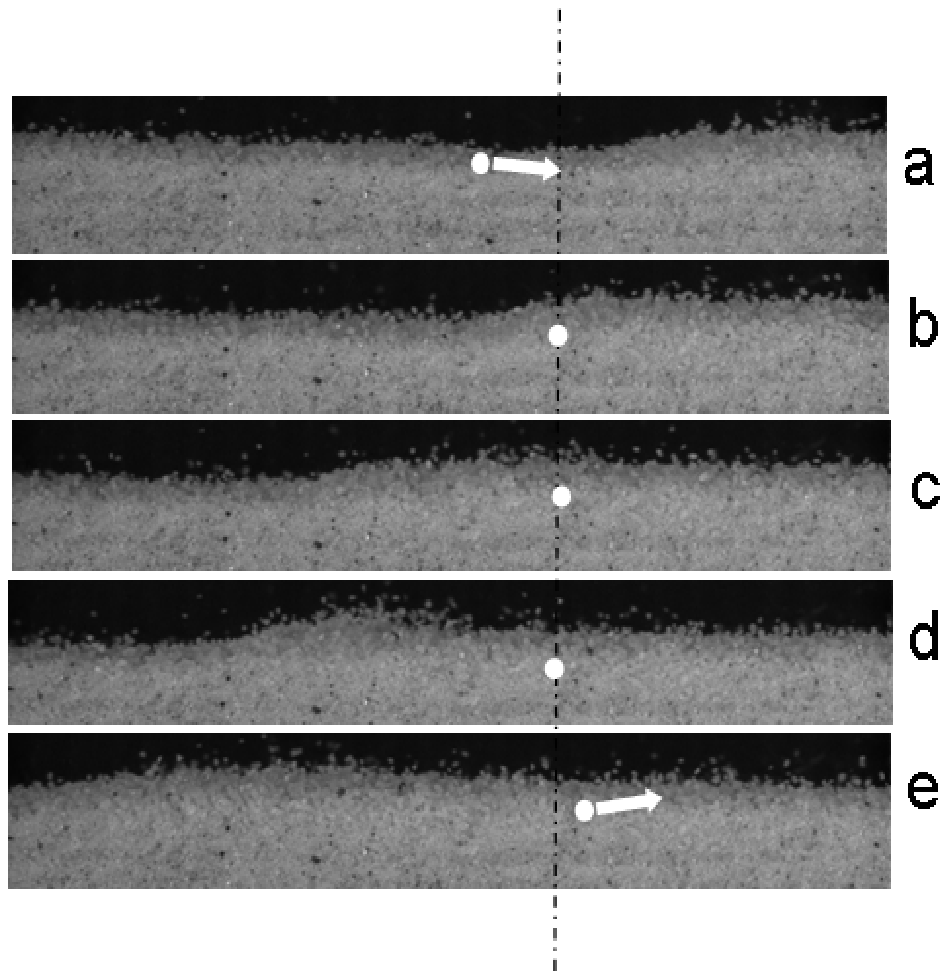}
\caption{\label{fig:f4} Sequence of pictures of an bump moving
uphill, with the same parameters described for Fig. 2(e). The
white circle represents a grain of sand (a) moving downhill as
part of the flowing layer, (b)-(d) trapped in the static layer as
the bump passes by from left to right, and (e) moving again as
part of the flowing layer. The horizontal length of the picture
corresponds to $22 mm$ in reality}
\end{figure}

All in all, the situation can be pictured as a solitary wave
crawling uphill underneath a shallow stream of sand flowing
downhill. In fact, the stop-and-go traffic model has been mapped
into the ``classical" Kortweg-de Vries (KdV) equation after a few
approximations, resulting in soliton-like solutions
\cite{Berg2001}. In the context of the KdV equation, the speed of
the soliton can be estimated as $v_s \sim \sqrt{gh_0}$ where $g$
and $h_0$ are the acceleration of gravity and the depth of the
unperturbed stream, respectively \cite{Toda1989}. If one assumes
the well accepted result that the depth of the flowing layer is
proportional to $F^{1/2}$ \cite{Midi2004}, we then get $v_s \sim
F^{1/4}$, which follows the experimental result shown in Fig. 3.

\begin{figure}
\includegraphics[height=1.4in, width=3.4in]{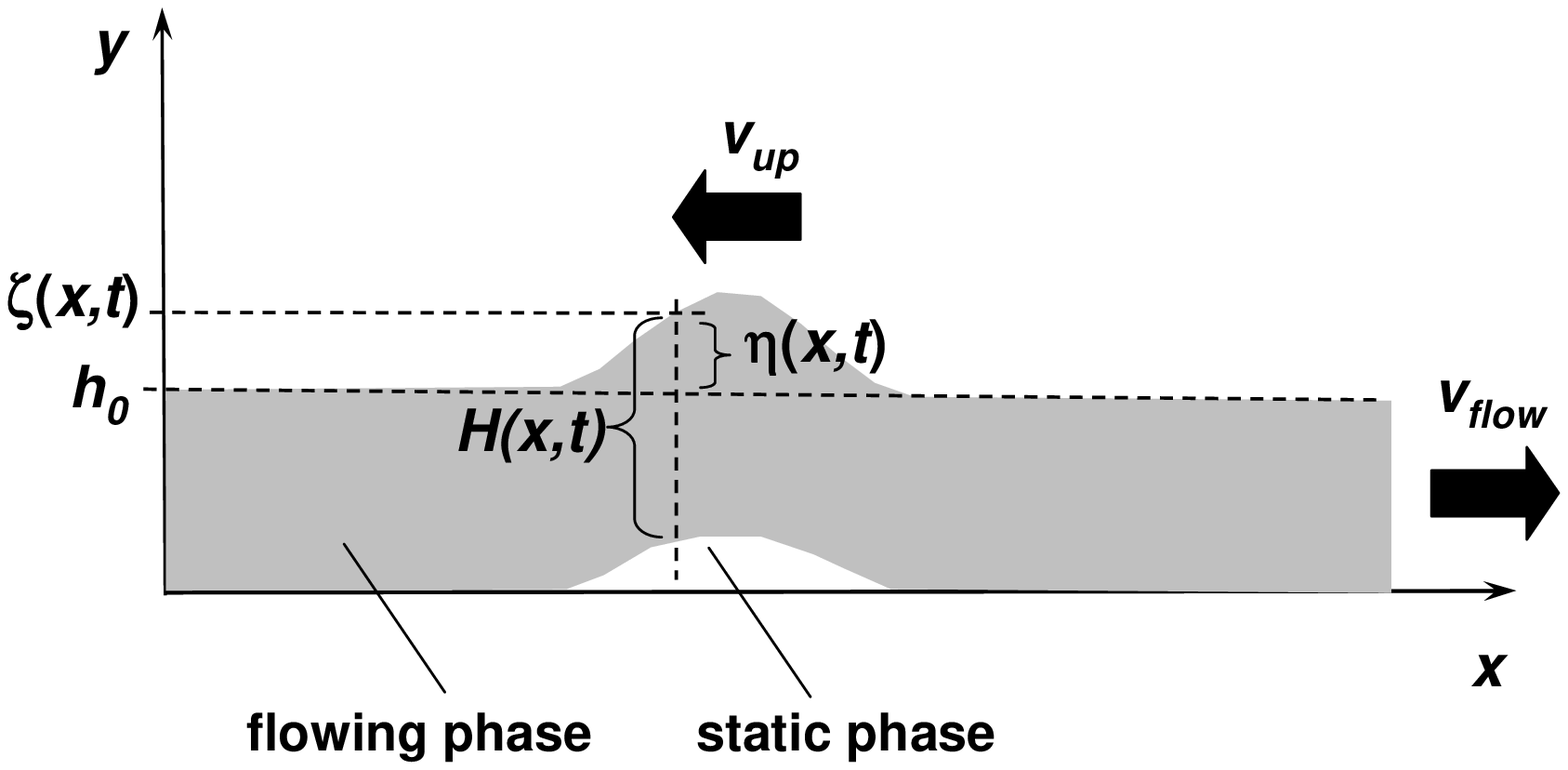}
\caption{\label{fig:f5} Diagram showing the neighborhood of a
``bump" which includes some parameters of the Saint-Venant model
described in the text.}
\end{figure}

Knowing that the KdV equation can be derived from mass and
momentum conservation equations applied to an incompressible fluid
flowing down a shallow channel \cite{Toda1989}, we could expect
that soliton solutions can be found if we describe our system
through hydrodynamics Saint-Venant equations modified to take into
account the particularities of granular matter
\cite{Douady1999,Andreotti2002} which we are writing relative to a
reference frame parallel to the average free surface of the heap
(see Fig. 5) . The first equation corresponds to mass
conservation:

\begin{equation}
\partial_{t} \zeta + \partial_x(\Gamma H^2/2)=0
\end{equation}

where $\zeta$ is the height of the free surface, $H$ is the
thickness of the flowing layer, and $\Gamma$ is the velocity
gradient transversal to the flowing layer, which is assumed
constant. Then, the momentum $q=\Gamma H^2 / 2$ evolves with $H$
due to the erosion/deposition process, like the one illustrated
for one grain in Fig. 4. The second equation corresponds to the
conservation of momentum, which can be seen as an equation
governing the evolution of the thickness of the flowing layer,
$H$:

\begin{equation}
\partial_{t} H + \partial_x(\Gamma H^2/2)= \frac{g}{\Gamma}(tan\theta-\mu(H))
\end{equation}

where $\mu(H)$ is the friction acting on the layer and
$tan\theta=-\partial_x\zeta$ is the slope of the free surface.

Let us assume that that the friction coefficient depends on $H$ in
the general form:

\begin{equation}
\mu = \alpha + (\beta + \frac{\gamma}{H})(h_0^2-H^2)\sqrt{\delta -
\varepsilon (h_0^2-H^2)}
\end{equation}

where the values of constants $\alpha$, $\beta$, $\gamma$,
$\delta$ and $\epsilon$ can be expressed in terms of experimental
parameters $v_{up}$, $v_{flow}$, $h_0$, $g$, and a characteristic
length $L_0$. Then, equations (1) and (2) can be transformed into
the ``textbook" KdV equation \cite{Toda1989}

\begin{equation}
L_0^2 v_{flow} \eta'' +6(v_{up}-v_{flow}) \eta
+30\frac{v_{flow}}{h_0} \eta^2=0
\end{equation}

In (4), $\eta = \zeta - {h_0}$ and $\eta'' = \frac{d^2 \eta}{d
\varphi^2}$, where $\varphi = x + v_{up}t$. One particular
solution of that equation, is

\begin{equation}
\zeta = h_0 + \frac{h_0}{5}sech^2(\frac{x + v_{up}t}{L_0})
\end{equation}

where we have taken $L_0 = 10 h_0$, $v_{flow} = \frac{1}{2} \Gamma
h_0 = \sqrt{gh_0}$, and $v_{up} = \frac{1}{3} v_{flow}$ in order
to match our experimental observations (i.e., mainly that the
uphill speed is roughly $1/3$ of the flow speed, and the height of
the perturbation is approximately $1/5$ of the thickness of the
unperturbed flowing layer). Formula (5) describes a ``textbook",
bell-shaped soliton moving upward. This solution holds for a
friction coefficient given by equation (3) which is a positive,
decreasing function of $H$ when reasonable experimental parameters
are introduced, in agreement with previous models
\cite{Andreotti2002}. Although we do not discard that
``non-textbook" solitonic equations might describe the observed
bumps without imposing such strong constraints, we stress that our
objective here is to show that ``textbook" solitons can be derived
if the system displays an appropriate $\mu(H)$.

We now present a number of additional experimental facts relative
to our uphill soliton waves. First, they have been observed in the
whole range of experimental parameters explored, i.e. $0.15 cm^3/s
\leq F \leq 3 cm^3/s$ and $0.3 cm \leq w \leq 3 cm$. As many
phenomena in granular matter, the uphill waves depend, to some
extent, on the way the heap is prepared, and show some degree of
``memory". When the heap is prepared from scratch, uphill bumps
nucleate at random places near the center of the heap, and, within
a few seconds, they typically appear near the open boundary. If
the experiment is stopped and re-started in such conditions,
solitons reappear immediately. That happens even if a layer of
sand less than $1 mm$-thick is removed from the surface. If a
thicker layer is removed, one has to wait a few seconds to observe
them. This is probably due to the formation of a ``compactified"
layer during the flow that is necessary for the formation of these
``jammed" bumps. Finally, all the results we have presented here
have been observed just for one type of sand that produces
sandpiles through ``revolving rivers" when dropped on a flat
horizontal surface \cite{Altshuler2003}. This fact supports the
idea that a quite specific $\mu(H)$ dependence is needed to
observe the solitary waves we report in this paper.


In summary, we have observed soliton-like instabilities in a flow
of sand established on a heap with open boundaries, moving against
the direction of the flow. The phenomenon can be understood in the
light of stop-and-go traffic arguments, even for this free surface
flow, and can be described by Saint-Venant equations adapted to
granular flows. The ``microscopic" mechanisms that make a certain
sand more suitable to show uphill bumps remains a mystery, but it
seems clear that the dependence of the friction coefficient of our
particular sand on the flowing depth is appropriate for the
appearance of solitons.

\begin{acknowledgments}

We thank {\O}. Johnsen and C. Noda for help in experiments and
image processing,  and  E. Cl\`{e}ment, A. Daerr, S. Franz, H.
Herrmann, J. Mar\'{\i}n, D. Mart\'{\i}nez, R. Mulet, D. Stariolo
and J.E. Wesfreid for useful discussions and comments. G. Quintero
and J. Fern{\'a}ndez collaborated in numerical calculations. We
appreciate financial support from the ``Abdus Salam" ICTP during
the last stage of this project.

\end{acknowledgments}



\begin{references}

\bibitem{Jaeger1996} H.Jaeger, S. R. Nagel and R. P. Behringer,
{\it Rev. Mod. Phys} {\bf 68}, 1259 (1996)

\bibitem{DeGennes1999} P. G. DeGennes, {\it Rev. Mod. Phys} {\bf 71}, S374 (1999)

\bibitem{Kadanoff1999} L. Kadanoff {\it Rev. Mod. Phys} {\bf 71}, 435 (1999)


\bibitem{BCRE1994} J. P. Bouchaud, M. E. Cates, J. Ravi Prakash and S. F. Edwards,
{\it J. Phys. I} {\bf 4}, 1383 (1994)

\bibitem{deGennes1995} P. G. de Gennes,
{\it C. R. Acad. Sci., Ser. IIb: Mech., Phys., Astron.} {\bf 321}, 501 (1995)

\bibitem{Aradian1999} A. Aradian, E. Raphael and P. G. de Gennes
{\it Phys. Rev. E.} {\bf 60}, 2009 (1999)

\bibitem{Douady1999} S. Douady, B. Andreotti and A. Daerr {\it Eur. Phys. J. B} {\bf
11}, 131 (1999)

\bibitem{Andreotti2002} B. Andreotti, A. Daerr, and S. Douady {\it Phys. Fluids}
{\bf 14}, 415 (2002).

\bibitem{Aranson2001} I. S. Aranson and L. S. Tsimring,
{\it Phys. Rev. E.} {\bf 64}, 020301 (2001)

\bibitem{Aranson2002} I. S. Aranson and L. S. Tsimring,
{\it Phys. Rev. E.} {\bf 65}, 061303 (2002)

\bibitem{Quartier2000} L. Quartier, B. Andreotti, S. Douady and A. Daerr
{\it Phys. Rev. E.} {\bf 62}, 8299 (2000)

\bibitem{Andreotti2001} B. Andreotti and S. Douady
{\it Phys. Rev. E.} {\bf 63}, 031305 (2001)

\bibitem{Pouliquen1997} O. Pouliquen, J. Delour and S. B. Savage
{\it Nature} {\bf 386}, 816 (1997)

\bibitem{Daerr1999} A. Daerr and S. Douady
{\it Nature} {\bf 399}, 241 (1999)

\bibitem{Forterre2001} Y. Forterre and O. Pouliquen
{\it Phys. Rev. Lett.} {\bf 86}, 5886 (2001)

\bibitem{Forterre2002} Y. Forterre and O. Pouliquen
{\it J. Fluid Mech.} {\bf 467}, 361 (2002)

\bibitem{Flekkoy2001} E. G. Flekk{\o}y, S. McNamara, K. J. M{\aa}l{\o}y
and D. Gendron {\it Phys. Rev. Lett.} {\bf 87}, 134302 (2001)

\bibitem{Gendron2001} D. Gendron, H. Troadic. K. J. M{\aa}l(\o)y
and E. Flekk{\o}y {\it Phys. Rev. E} {\bf 64}, 021509 (2001).

\bibitem{Altshuler2003} E. Altshuler, O. Ramos, E. Mart\'{\i}nez,
A. J. Batista-Leyva, A. Rivera, and K. E. Bassler, {\it Phys. Rev.
Lett.} {\bf 91}, 014501 (2003).

\bibitem{Douady2005} S. Douady, A. Manning, P. Hersen, H. Elbelrhiti,
S. Protiere, A. Daerr, B. Kabbachi, arxiv.org/abs/nlin/0412047v1

\bibitem{Midi2004} G. D. R. Midi, {\it Eur. Phys. J. E} {\bf 14}, 341
(2004)

\bibitem{Helbing2001} D. Helbing, {\it Rev. Mod. Phys.} {\bf 73}, 1067 (2001).

\bibitem{Berg2001} P. Berg, A. Woods, {\it Phys. Rev.
E.} {\bf 64}, 035602 (2001).

\bibitem{Toda1989} M. Toda, {\it Nonlinear Waves and Solitons},
Kluwer Academic Publishers, Dordrecht, 1989



\end{references}

\end{document}